\documentclass[fleqn,twoside]{article}
\usepackage[headings]{espcrc2}

\readRCS
$Id: espcrc2.tex,v 1.2 2004/02/24 11:22:11 spepping Exp $
\usepackage{graphicx}
\usepackage[figuresright]{rotating}

\newcommand{\AmS}{{\protect\the\textfont2
  A\kern-.1667em\lower.5ex\hbox{M}\kern-.125emS}}

\newcommand{\Mvec}{\mbox{\rm\bf M}}
\newcommand{\Li}{\mbox{Li}}

\title{
{\footnotesize DESY 05-003, SFB/CPP-06-06\hfill hep-ph/xxyymmm
}\\
$O(\alpha_s^3)$ contributions to $F_L^{Q\overline{Q}}(x,Q^2)$ for 
large virtualities}

\author{Johannes Bl\"umlein\address[MCSD]{Deutsches 
Elektronen-Synchrotron, DESY, Platanenallee 6, D-15738 Zeuthen, 
Germany}%
\thanks{This work was supported in part by DFG Sonderforschungsbereich
Transregio 9, Computergest\"utzte Theoretische Physik.}}
       
\runtitle{$O(\alpha_s^3)$ contributions to $F_L^{Q\overline{Q}}(x,Q^2)$}
\runauthor{J. Bl\"umlein}

\begin{document}

\begin{abstract}
\noindent
The $O(\alpha_s^3)$ contributions to the heavy flavor Wilson coefficients for the 
structure function $F_L(x,Q^2)$ are calculated in the region $Q^2 \gg m^2$ using
the renormalization group method.
\vspace{1pc}
\end{abstract}

\maketitle
\section{INTRODUCTION}

\vspace{1mm}
\noindent
The heavy flavor contributions to unpolarized deep-inelastic structure functions 
are large in the region of small values of $x$, see e.g. \cite{BR}. At present they 
are known 
to $O(\alpha_s^2)$ \cite{HF2}, while the anomalous dimensions and Wilson 
coefficients for the light parton contributions were calculated to $O(\alpha_s^3)$
\cite{THRL,MVV3}. Since the scaling violations of the light parton and heavy flavor 
terms
are different, the knowledge of both contributions at the same order of precision
is highly desirable to perform QCD analyses. While a complete calculation 
of the $O(\alpha_s^3)$ heavy flavor 
Wilson coefficients is still a technical problem, the 
calculation of
these quantities in the asymptotic regime $Q^2 \gg m^2$ seems feasable. 
Calculations of this kind were carried out at the $O(\alpha_s^2)$ level before  
\cite{BMSMN,ASYMP1}. 

In the region of smaller values of $x$ the structure function 
$F_L(x,Q^2)$ can be used to put stringent constraints on the gluon 
distribution \cite{FELT}. It is therefore importnat to calculate this 
quantity as precisely as possible. In this paper we give a brief outline 
of the calculation of the heavy flavor Wilson coefficients for the 
longitudinal structure function $F_L(x,Q^2)$ in the asymptotic region. 
Details are presented in \cite{BFMNK}. 
\section{THE METHOD}

\vspace{1mm}
\noindent
The twist--2 contributions to structure functions in deeply inelastic 
scattering due to massless partons are described as convolutions between
the (massless) parton densities and the Wilson coefficients due to mass 
factorization. Since the scale evolution of parton densities is free of 
quark mass effects, the heavy flavor contributions to the structure 
functions result form the associated Wilson coefficients only. In the 
limit 
$Q^2 \gg m^2$ the non--power contributions to these coefficient functions    
obey the factorization relation
\begin{eqnarray}
\label{eq1}
\lefteqn{H_{I,l}^{{\rm S,NS},g} \left(\frac{Q^2}{\mu^2}, \frac{m^2}{\mu^2}\right)}
\nonumber\\ 
&=&
A_{k,l}^{{\rm S,NS},g} \left( \frac{m^2}{\mu^2}\right) 
\otimes C_{I,k}^{{\rm S,NS},g} \left(\frac{Q^2}{\mu^2}\right), 
\end{eqnarray}
with $\mu$ the factorization scale, $A_{k,l}$ partonic operator matrix 
elements and $C_{I,k}$ the respective light-parton Wilson coefficients. 
The operator-matrix elements are process-independent quantities.
Eq. (\ref{eq1}) allows to calculate all logarithmic contributions and the constant 
term in the on-mass-shell scheme. For the $O(\alpha_s^3)$ asymptotic heavy flavor 
Wilson coefficients for $F_2(x,Q^2)$ the 3-loop operator matrix elements are 
required. In case of $F_L(x,Q^2)$ the $O(\alpha_s^3)$ contributions depend on the 
2-loop operator matrix elements and the massless Wilson coefficients at 
$O(\alpha_s^3)$ \cite{MVV3} only, since at leading order the Wilson coefficient is 
scale-independent.
\section{HEAVY FLAVOR WILSON COEFFICIENT FOR \boldmath{$F_L(x,Q^2)$}}

\vspace{1mm}
\noindent
To $O(\alpha_s^3)$ three heavy flavor Wilson coefficients contribute: 
$H_{L,g}^{\rm S}, H_{L,q}^{\rm PS},H_{L,q}^{\rm NS}$. In the asymptotic region
$Q^2 \gg m^2$ they are given in terms of Mellin convolutions between the
light-parton Wilson coefficients and the corresponding operator matrix elements
$A_{i,j}^{(k)}$~:
{\small
\begin{eqnarray}
\label{eq1a}
\lefteqn{H_{L,g}^{\rm S}\left(\frac{Q^2}{m^2}, \frac{m^2}{\mu^2}\right)
= a_s    \widehat{C}^{(1)}_{L,g}\left(\frac{Q^2}{\mu^2}\right)}\nonumber\\
&& +  a_s^2  \left[
A_{Q,g}^{(1)}\left(\frac{\mu^2}{m^2}\right) \otimes
                  C^{(1)}_{L,q}\left(\frac{Q^2}{\mu^2}\right)
 +                \widehat{C}^{(2)}_{L,g}\left(\frac{Q^2}{\mu^2}\right)\right]
\nonumber\\
&& +a_s^3 \left[
A_{Q,g}^{(2)}\left(\frac{\mu^2}{m^2}\right) \otimes 
C^{(1)}_{L,q}\left(\frac{Q^2}{\mu^2}\right)
+ A_{Q,g}^{(1)}\left(\frac{\mu^2}{m^2}\right) \right. \nonumber\\ && \left. 
\hspace*{8mm} \otimes 
                  C^{(2)}_{L,q}\left(\frac{Q^2}{\mu^2}\right)
 + \widehat{C}^{(3)}_{L,g}\left(\frac{Q^2}{\mu^2}\right)\right]
\nonumber\\ 
\label{eq1b}
\lefteqn{H_{L,q}^{\rm PS}\left(\frac{Q^2}{m^2}, \frac{m^2}{\mu^2}\right)
= a_s^2  \widehat{C}^{{\rm PS},(2)}_{L,q}\left(\frac{Q^2}{\mu^2}\right)}
\nonumber\\ &&
 +  a_s^3  \left[
A_{Qq}^{{\rm PS},(2)}\left(\frac{\mu^2}{m^2}\right) \right.
\nonumber\\ && \left. \hspace*{7mm}
\otimes C^{(1)}_{L,q}\left(\frac{Q^2}{\mu^2}\right)
 +\widehat{C}^{{\rm PS},(3)}_{L,q}\left(\frac{Q^2}{\mu^2}\right)\right]
\nonumber\\
\lefteqn{H_{L,q}^{\rm NS}\left(\frac{Q^2}{m^2}, \frac{m^2}{\mu^2}\right)
= a_s^2    \widehat{C}^{{\rm NS},(2)}_{L,q}\left(\frac{Q^2}{\mu^2}\right)} 
\nonumber\\ &&
 +  a_s^3  \left[
A_{qq,Q}^{{\rm NS},(2)}\left(\frac{\mu^2}{m^2}\right) \otimes
                  C^{(1)}_{L,q}\left(\frac{Q^2}{\mu^2}\right) \right.
\nonumber\\ && \left. +
\widehat{C}^{{\rm NS},(3)}_{L,q}\left(\frac{Q^2}{\mu^2}\right)\right]~,
\nonumber
\end{eqnarray}
}

\noindent
with $\hat{f} = f(N_F+1) - f(N_F)$.
The operator matrix elements have the form
\begin{eqnarray}
A_{i,j}^{(k)}\left(\frac{m^2}{\mu^2}\right) = \sum_{l=1}^{k} \hat{a}_{i,j}^{l,(k)}
\ln^l\left(\frac{m^2}{\mu^2}\right) + a_{i,j}^{(k)}~.
\nonumber
\end{eqnarray}
They were calculated in \cite{BMSMN} up to $k=2$ in $z$-space. 
The operator matrix elements take their most simple structure in Mellin space.
As an example, the coefficient $a_{Qg}^{(2)}(N)$ reads
{\small
\begin{eqnarray}
\lefteqn{a_{Qg}^{(2)}(N) = 4 C_F T_R \Biggl\{
                   \frac {{N}^{2}+N+2}{N \left( N+1 \right)  \left( N+2 \right) }
                   \Biggl[}\nonumber\\
&&                   - \frac{1}{3} S_1^3(N-1) + \frac{4}{3} S_3(N-1)
                   - S_1(N-1) 
\nonumber\\ && \times
S_2(N-1) - 2 \zeta_2 S_1(N-1) \Biggr]
\nonumber\\ && 
+ \frac{2}{N(N+1)} S_1^2(N-1)
\nonumber\\ && 
+  \frac{N^{4} +16\,{N}^{3} +15\,{N}^{2}-8\,N-4}
        {N^2 \left( N+1 \right)^{2} \left( N+2 \right) } S_2(N-1)
\nonumber\\ & & 
+\frac { 3\,{N}^{4}+2\,{N}^{3}+3\,{N}^{2}-4\,N-4}
      {2 N^2 \left( N+1 \right) ^{2} \left( N+2 \right) } \zeta_2
\nonumber\\ & &
+\frac {N^4-N^3-16 N^2 + 2N +4}
       {N^2 \left( N+1 \right) ^{2}\left( N+2 \right)} S_1(N-1)
\nonumber\\ & &
+ \frac {P_2(N)}{2 N^4 \left( N+1 \right) ^{4} \left( N+2 \right) }\Biggr\}
\nonumber\\
&& +4 C_A T_R\Biggl\{ \frac{N^2+N+2}{N(N+1)(N+2)}
\Biggl[ 4 \Mvec\left[\frac{\Li_2(x)}{1+x}\right](N) 
\nonumber\\ &&
+\frac{1}{3}
S_1^3(N)+3 S_2(N) S_1(N) + \frac{8}{3} S_3(N)
\nonumber\\& & 
+ \beta''(N+1) -
 4 \beta'(N+1) S_1(N) - 4 \beta(N+1)
\zeta_2
\nonumber\\& & 
+\zeta_3\Biggr]
- \frac {{N}^{3}+8\,{N}^{2}+11\,N+2}{N \left( N+1 \right) ^{2} \left( N+2
 \right) ^{2}} S_1^2(N)
\nonumber\\ & &
-2\,{\frac {N^4 - 2 N^3 + 5 N^2+ 2 N + 2}
{ \left( N-1 \right)  N^2 \left( N+1 \right) ^{2} \left( N+2
\right) }}
\zeta_2\nonumber\\
& & 
- \frac {P_3(N)}
           { (N-1) N^2 \left( N+1 \right) ^{2} \left( N+2 \right) ^{2}}
           S_2(N)
\nonumber\\
& & 
- \frac {P_4(N)}
{ N \left( N+1 \right) ^
{3} \left( N+2 \right) ^{3} } S_1(N)
\nonumber\\
& & 
-4 \, \frac { \left( {N}^{2} - N -4 \right)}
            { \left( N+1 \right) ^{2} \left( N+2 \right) ^{2}} \beta'(N+1)
\nonumber\\ &&
+ \frac{P_5(N)}{(N-1) N^4 (N+1)^4 (N+2)^4}\Biggr\}~,
\nonumber
\end{eqnarray}
}
\noindent
where
\begin{eqnarray}
P_2(N) &=&  12 N^{8}
             +54 N^{7}
            +136 N^{6}
            +218 N^{5} \nonumber\\ &&
            +221 N^{4}
            +110 N^{3}
              -3 N^{2}
             -24 N -4
     \nonumber\\
P_3(N) &=& 7 N^5 + 21 N^4 + 13 N^3 + 21 N^2 +18 N 
\nonumber\\ &&+16     \nonumber\\
P_4(N) &=& {N}^{6}+8\,{N}^{5}+23\,{N}^{4}+54\,{N}^{3}+94\,{N}^{2}
\nonumber\\ 
&&
+72\,N+8\nonumber\\
P_5(N) &=&      2\,{N}^{12}
      +20\,{N}^{11}
      +86\,{N}^{10}
     +192\,{N}^{9}
\nonumber\\ 
&&
     +199\,{N}^{8}
          -{N}^{7}
     -297\,{N}^{6}
     -495\,{N}^{5}
\nonumber\\ 
&&
-514\,{N}^{4} 
     -488\,{N}^{3}
     -416\,{N}^{2}
      -176\,N
\nonumber\\ & &
       -32
\nonumber\\
\beta(N) &=& \frac{1}{2} \left[\psi\left(\frac{N+1}{2}\right) - \psi\left(\frac{N}{2}
\right)\right]~. \nonumber
\end{eqnarray}
Similar, even simpler, expressions are obtained for the other matrix 
elements
\cite{BFMNK}. The non-logarithmic contributions to 
$H_{L,g(q)}^{\rm S,PS,NS}$ contain 
the respective
light-parton Wilson coefficients of given order and a term, which 
consists out of 
convolutions of operator matrix elements, splitting functions and lower order
Wilson coefficients.  

Since 
\begin{eqnarray}
\lefteqn{\Mvec\left[\frac{\Li_2(x)}{1+x}\right](N) - \zeta_2 \beta(N)}\nonumber\\ 
&&= (-1)^N\left[S_{-2,1}(N)(N-1) + \frac{5}{8} \zeta_3\right]
\end{eqnarray}
also the heavy flavor Wilson coefficients to $O(\alpha_s^3)$ belong to the
class which can be represented by harmonic sums {\it without} an index $k 
= -1$, as observed earlier	
for all known 2--loop Wilson coefficients \cite{STR2} and the 3--loop anomalous 
dimensions \cite{STR31,STR32}. Due to this the general class of harmonic sums 
\cite{HSUM} is drastically reduced \cite{STR31,HLHC} even before referring to 
structural relations.

\section{SMALL \boldmath{$x$} LIMIT}

\vspace{1mm}
\noindent
In the small-$x$ limit the asymptotic heavy flavor Wilson coefficient
for $F_L^S(x,Q^2)$ obey 
\begin{eqnarray}
\lefteqn{H_{L}^{S}(z) \propto~a_s^2~\frac{d_1^{(1)}}{z} + 
\sum_{k=2}^\infty a_s^{k+1}} \nonumber\\
&\times& 
\left[
d_k^{(1)} \frac{\ln^{k-1}(z)}{z} +
d_k^{(2)} \frac{\ln^{k-2}(z)}{z} + \ldots \right]~.
\end{eqnarray}
for $Q^2 = \mu^2$. The lowest order result $d_{1,i}^{(1)} = -32 C_i T_R/9,~i=A,F$
\cite{CCH} agrees with the corresponding limit of the complete calculation
\cite{BMSMN}. At $O(\alpha_s^3)$ the leading terms are due to the massless
contributions \cite{MVV3} in the $\overline{\rm MS}$ scheme, while the 
small $x$ heavy flavor contributions emerge in $d_k^{(2)}$ for the first time\footnote{
We corrected typographical errors contained in Nucl. Phys. (Proc. Suppl.) 
{\bf 157} (2006) 2.}  
\begin{eqnarray}
\lefteqn{d_{2, i}^{(2)} = -32 C_i C_F T_R}\nonumber\\
\hspace*{-2mm} & &\times
\left[ \frac{1}{3}\ln^2\left(\frac{Q^2}{m^2}\right)
- \frac{10}{9} \ln\left(\frac{Q^2}{m^2}\right) +
\frac{28}{27}\right] + \tilde{d}_{2,i}^{(2)} \nonumber\\
& &-\delta_{iA} \frac{256}{27} C_F T_R^2(2N_f+1) \ln\left(\frac{Q^2}{m^2}\right) 
\end{eqnarray}
Here, $\tilde{d}_{2,i}^{(2)}$ are contributions due to the massless Wilson 
coefficient. While the genuine heavy flavor terms in the sub-leading order 
scale with the color factor, the $\tilde{d}_{2,i}^{(2)}$-terms do not.
\section{CONCLUSIONS}

\vspace{1mm}
\noindent
In the region $Q^2 \gg m^2$ the 3--loop heavy flavor Wilson coefficients 
for the structure function $F_L(x,Q^2)$ can be calculated using the 
renormalization group method. They depend on the massless 
$\overline{\rm MS}$ scheme Wilson coefficients up to 3--loop order and
massive operator matrix elements and anomalous dimensions up to 2--loop 
order. The complexity of the structure of the Wilson coefficients beyond 
the $\overline{\rm MS}$ light--parton 3--loop contributions is that of
the 2--loop anomalous dimensions. Their representation in Mellin space
shows that indices $k=-1$ do not occur in the respective harmonic sums, as 
observed for a large variety of other cases before. The pure 
heavy flavor terms at $O(\alpha_s^3)$ contribute at one order less than 
the leading logarithmic order in the small $x$ region.   

\vspace{1mm}\noindent
{\bf Acknowledgment.}~~I would like to thank the organizers of the conference 
for arranging a very interesting meeting in a splendid atmosphere.

\end{document}